\documentclass[10pt,letterpaper]{article}

\usepackage{./style/cogsci}
\cogscifinalcopy 
\usepackage{subfig}
\usepackage{graphicx}
\graphicspath{{./figs/}}
\usepackage{pslatex}
\usepackage{apacite}
\usepackage{float} 

\usepackage{xcolor,soul}

\title{Extracting low-dimensional psychological\\representations from convolutional neural networks}
 
\author{{\bf Aditi Jha$^1$ (aditijha@princeton.edu)}\\
{\bf Joshua Peterson$^2$ (joshuacp@princeton.edu)}\\
{\bf Thomas L. Griffiths$^{2,3}$ (tomg@princeton.edu)}\\
$^1$Department of Electrical Engineering\\
$^2$Department of Computer Science\\
$^3$Department of Psychology\\
Princeton University}

\begin{document}

\maketitle

\begin{abstract}
Deep neural networks are increasingly being used in cognitive modeling as a means of deriving representations for complex stimuli such as images. While the predictive power of these networks is high, it is often not clear whether they also offer useful explanations of the task at hand. Convolutional neural network representations have been shown to be predictive of human similarity judgments for images after appropriate adaptation. However, these high-dimensional representations are difficult to interpret. Here we present a method for reducing these representations to a low-dimensional space which is still predictive of similarity judgments. We show that these low-dimensional representations also provide insightful explanations of factors underlying human similarity judgments. 

\end{abstract}

\section{Introduction}
Judging similarity between any pair of stimuli is an ambiguous problem: deciding what counts as similar is subjective and sensitive to context \cite{Medin1993}.
Nevertheless, people are relatively consistent in making similarity judgments, which is perhaps explained in part by the biases they develop towards emphasizing some stimulus features over others (\textit{e.g.,} shape over size, color, or material; \citeNP{Diesendruck2003}). Understanding the features (and the weights upon them) that people employ when evaluating the similarity of complex stimuli like images remains an open problem.

Deep neural networks have been demonstrated to be predictive of multiple aspects of human visual perception in visuoperceptual tasks  \cite<\textit{e.g.,}>{Lake2015,Kubilius2016}. This utility has led to their increasing use as proxies of human cognition to understand mechanisms underlying cognitive processes or as proofs-of-concept to establish the possibility of a certain cognitive strategy \cite{Kriegeskorte2015, Cichy2019}. For example, \citeA{Sanders2020} show that CNNs can be trained using multidimensional scaling representations to derive psychological representations of images. In other work, \citeA{Peterson2017} show correspondences between similarities in convolutional neural net (CNN) representations and human similarity judgments for natural images. They find that, while out-of-the-box CNN representations are only partially reflective of human psychological representations, they can be adapted to support a more fine-grained correspondence. 


The success of CNNs in predicting human similarity judgments suggests that they might also be helpful in identifying the features which inform those judgments. However, CNN representations are high-dimensional, potentially redundant, and likely include psychologically irrelevant stimulus information. An important question, given their increasing use in cognitive modeling is how many relevant features/dimensions they really contribute and what the nature of those features might be.

In this work, we propose a method inspired by previous work by \citeA{Rumelhart1993} which we call similarity-driven dimensionality reduction (SimDR), which obtains low-dimensional projections of CNN image representations that best capture human similarity judgments. Surprisingly, our method reveals that human similarity judgments continue to be well-preserved even up to two orders of magnitude fewer dimensions than previous work. This suggests that the dimensionality of psychological representations is considerably less than the full set of CNN features. We further probe the individual dimensions of these representations that capture concepts essential to judging similarity, and find that most of them are interpretable. In particular, we show that broad categories are given more importance by our model than finer ones captured in subsequent dimensions, in line with the hierarchical structure oft-found to characterize human cognition \cite{Cohen2000, Rogers2004SemanticCA}.

\section{Method}
\citeA{Peterson2017} show that the final representation layer of a CNN can be adapted to better predict human similarity judgments. The size of the final representation in CNNs is typically of the order of $10^3$, which makes interpretation difficult.
To serve our purpose of leveraging CNN representations to understand human similarity judgments, we require representations that are interpretable and can give us insights into the actual cognitive task.

\citeA{Rumelhart1993} constructed a connectionist model to mimic human similarity judgments. The model takes two stimuli as input and outputs a similarity judgment. The hidden layer is of lower dimensionality than the input, resulting in a compressed representation.
Extending this idea to modern CNNs, our method (SimDR) reduces the CNN representations of images to a low-dimensional space which is optimal for predicting human similarity judgments. 
If the obtained representations have sufficiently low dimensionality, we can interpret individual dimensions to see what they capture and make inferences about similarity judgments in humans. This model and the data used are explained in the following sections.

\subsection{Datasets}
\citeA{Peterson2017} collected six human similarity datasets for natural images drawn from the following domains: animals, vehicles, vegetables, fruits, furniture and a dataset encompassing a variety of domains (``various''). Each of these sets comprises pairwise similarity ratings from ten people for 120 images, which we employ in all analyses that follow.

\subsection{Similarity-driven Dimensionality Reduction}
\citeA{Peterson2017} showed that the final, fully-connected representation layer of VGG-19 \cite{Simonyan15} is most predictive of human similarity judgments, hence we use the same 4096-dimensional VGG-19 representations for all our experiments. The task of obtaining low-dimensional representations of images which capture factors underlying human similarity judgments is split by SimDR into two objectives: (a) projecting VGG-19 representations to a low-dimensional space, and (b) predicting human similarity judgments using the low-dimensional representations.

\begin{figure}[!t]
\begin{center}
\includegraphics[width=0.9\linewidth, trim=3mm 3mm 3mm 4mm, clip=true]{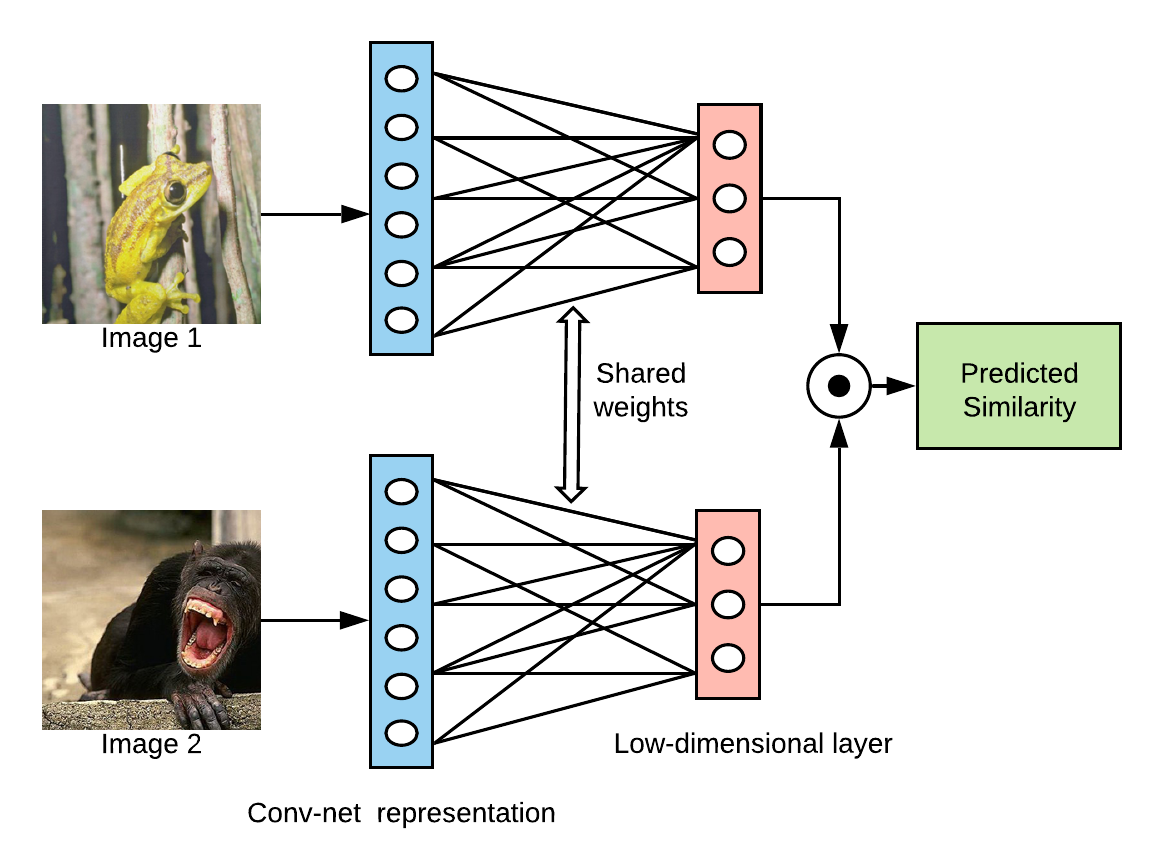}
\end{center}
\caption{Overview of SimDR. CNN representations for an image pair are down-projected using a shared low-dimensional bottleneck layer. An inner product of the outputs gives predicted similarity rating for the input pair.}
\label{fig::model_diag}
\end{figure}

These two objectives are jointly optimized leading to low-dimensional representations that are predictive of human similarity judgments. VGG-19 representations of two input images are passed through a single linear layer of small width (\textit{i.e.,} a bottleneck layer) which projects them to a lower-dimensional space. This is followed by an inner product of the outputs of the bottleneck layer to obtain the predicted similarity rating for the input pair (Fig.~\ref{fig::model_diag}). The inner product is our representational similarity measure, which contrasts with \citeA{Rumelhart1993}, and more directly generalizes the method of \citeA{Peterson2017}. For both input images, the weights of the bottleneck layer are shared. The weights are learned by back-propagating the loss incurred during the prediction of human similarity judgments, hence optimizing the projected representations to predict human similarity judgments. This contrasts with the method of \citeA{Peterson2017}, which learns weights for each of the 4096 input dimensions, or principal component analysis (PCA), which preserves as much information as possible as opposed to just that which is relevant to human similarity judgments (and thus may inflate the estimated intrinsic dimensionality).

We first trained a separate model for each dataset. CNN feature vectors were first normalized such that their norms were one. We used mean squared error loss with L2 regularization to train each model. The L2 coefficient was selected between $10^{-3}$ and $10^3$ by $6$-fold cross-validation over the $120$ images. Further, for every dataset, the number of nodes in the bottleneck layer is varied in the range of $1-64$. We also compare the above with a simple unsupervised baseline that alternatively obtains low-dimensional representations by running PCA over the input VGG-19 representations. These low-dimensional representations are then transformed by ridge regression using the method of \citeA{Peterson2017} to predict similarity ratings. As above, we vary the number of principal components in the range of $1-64$.

\begin{table}[!b]
\begin{center} 
\setlength{\tabcolsep}{3.5pt}
\begin{tabular}{lcccc}
\hline
Dataset & Raw & \citeA{Peterson2017} & SimDR & PCA \\
\hline
Animals & 0.58 & 0.74 & 0.64 & 0.47\\
Vehicles & 0.51 & 0.58 & 0.57 &0.51  \\
Fruits & 0.27 & 0.36 & 0.30 & 0.27\\
Furniture & 0.19 & 0.35 & 0.33 & 0.28\\
Various & 0.37 & 0.54 & 0.50 & 0.31\\
Vegetables & 0.27 & 0.34 & 0.30 & 0.32\\
\hline
\end{tabular} 
\caption{$R^2$ scores for all datasets (SimDR values are for bottleneck layer of size 64).} 
\label{table::r2score}
\end{center} 
\end{table}

\begin{figure*}[!t]
\begin{center}
\includegraphics[width=1.0\linewidth, trim=2mm 2mm 3mm 3mm, clip=true]{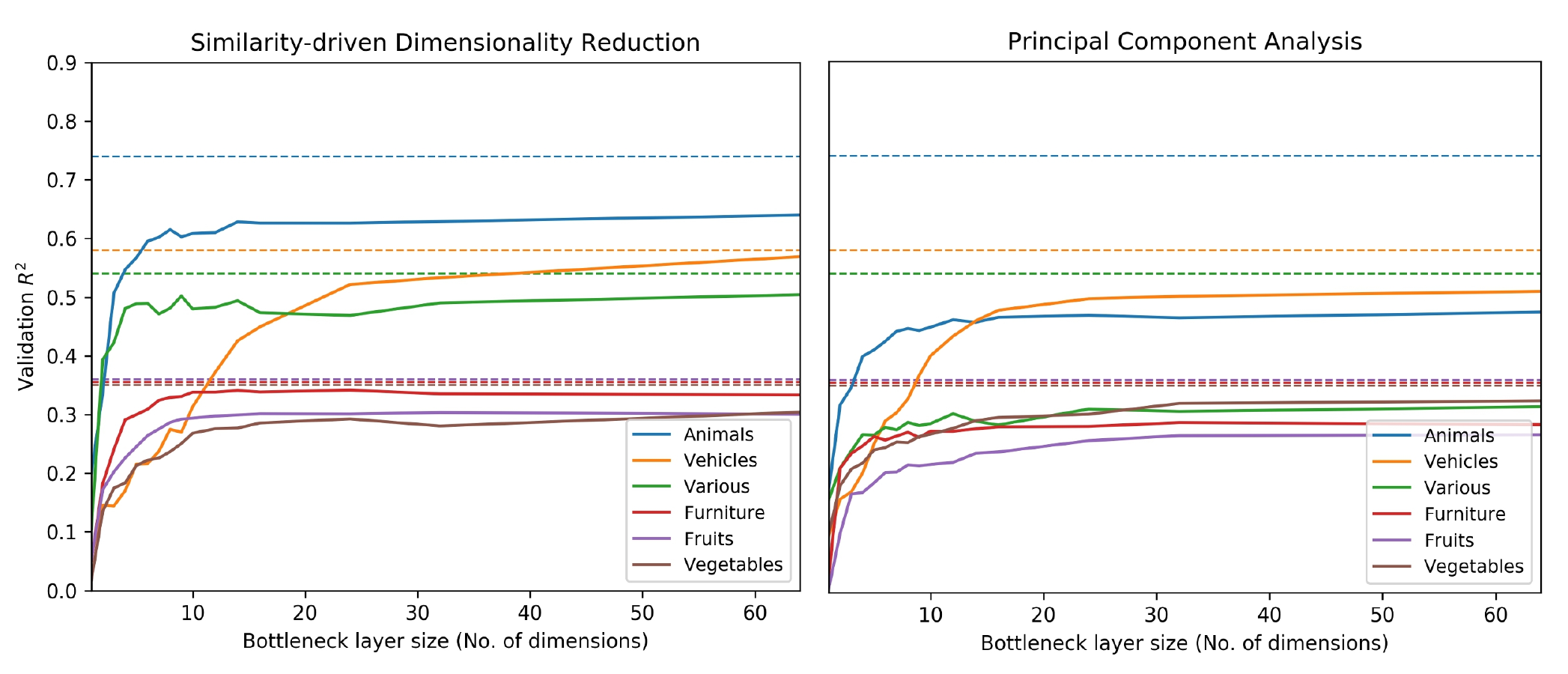}
\end{center}
\caption{Explained variance ($R^2$) of our models in predicting human similarity judgments on each dataset. The dashed lines correspond to the prediction performance in \citeA{Peterson2017} when all input dimensions are used.} 
\label{fig::dimvsr2}
\end{figure*}

\section{Few dimensions predict similarity judgments}
\begin{figure}[!b]
\begin{center}
\includegraphics[width=1.0\linewidth, trim=7mm 2mm 13mm 10mm, clip=true]{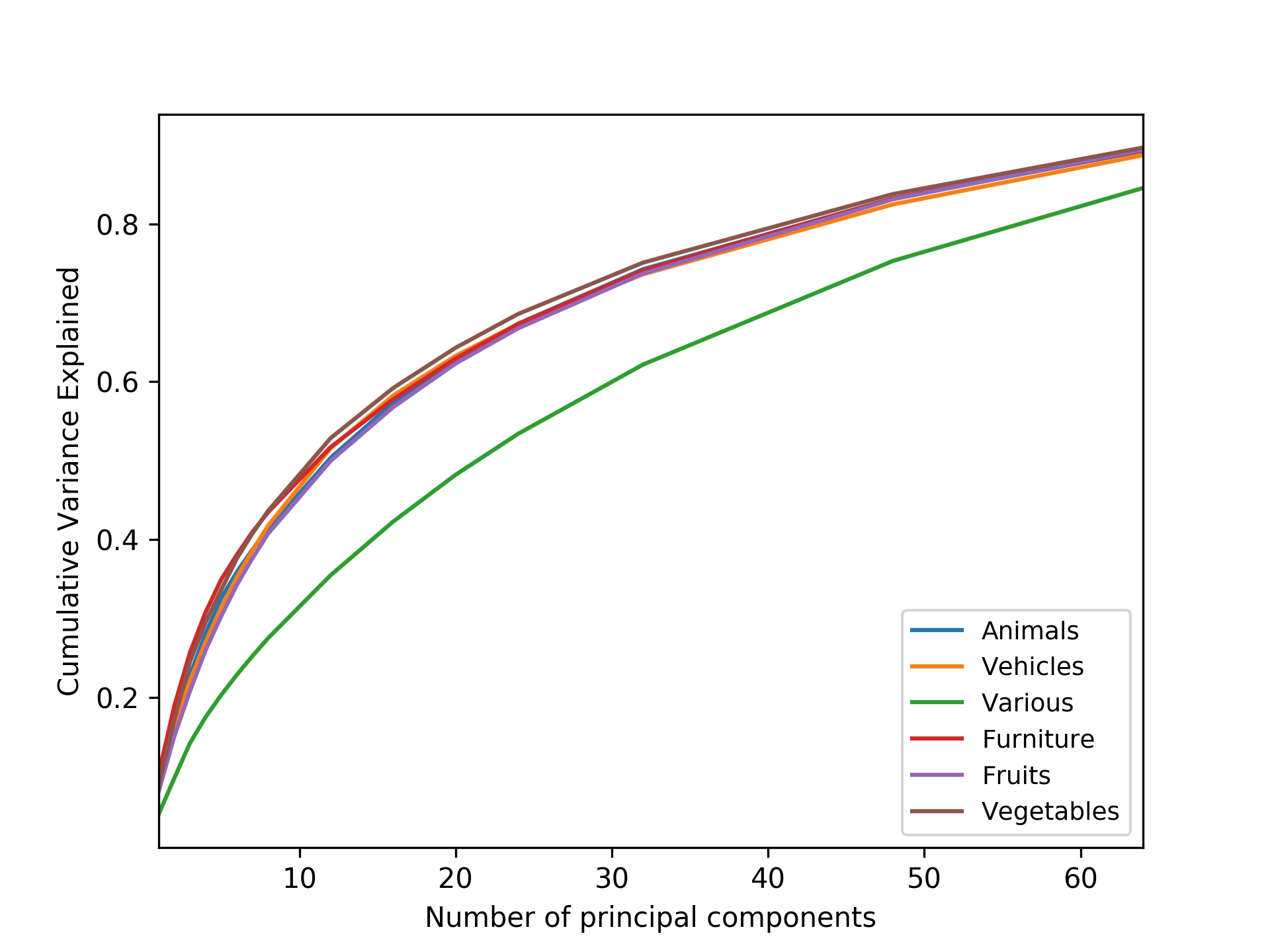}
\end{center}
\caption{Cumulative variance explained in the full VGG-19 representations as a function of principal component.}
\label{fig::pca_data}
\end{figure}

We observe for all datasets that the SimDR $R^2$ score at 64 dimensions is higher than that of the raw (untransformed) CNN representations (Table~\ref{table::r2score}). The PCA-based model performed worse than SimDR for all datasets (except for the \textit{vegetables} dataset), suggesting that supervision is much more selective of the human-relevant dimensions. We also observe that the prediction performance of SimDR quickly saturates as the number of dimensions increases beyond $10-20$, approaching the prediction performance obtained using all VGG-19 features (Fig.~\ref{fig::dimvsr2}; dashed lines). Notably, the \textit{animals} dataset requires only 6 nodes to achieve an $R^2$ score of 0.6 while the \textit{various} dataset achieves an $R^2$ of 0.49 at 6 nodes. These results strongly suggest that human similarity judgments can be captured by considerably fewer dimensions (by at least two orders of magnitude) than those comprising VGG-19 representations, and more generally that psychological representations as measured by similarity experiments are much lower-dimensional than CNN representations. Additional evidence for this can be seen in the intrinsic dimensionality of the CNN representations themselves without respect to human judgements. Fig.~\ref{fig::pca_data} illustrates this using PCA: cumulative variance explained is shown as a function of the number of components, for each dataset. Notably, the dimensionality elbow is both longer and later than those in Fig.~\ref{fig::dimvsr2}. Interestingly, CNNs also appear to assign equal dimensionality to all datasets (except \textit{various}), apparently much unlike humans (Fig.~\ref{fig::dimvsr2}).

\section{Interpretation of low-dimensional features}
Now that we have demonstrated the sufficiency of low-dimensional representations to predict similarity judgments, we can attempt to interpret the reduced dimensions. For this experiment, we focus on the top 3 datasets based on $R^2$ score---\textit{animals}, \textit{vehicles}, \textit{various}. As mentioned above, SimDR achieves an $R^2$ score of 0.6 on the \textit{animals} dataset  using only 6 dimensions. On the \textit{various} dataset, it achieves an $R^2$ score of 0.49 using 6 dimensions, and an $R^2$ score of 0.45 on \textit{vehicles} dataset  using 16 dimensions. We fix these as the bottleneck layer sizes for each of these datasets. The aforementioned dimensions for each of the three datasets are chosen by visually identifying an elbow in performance (Fig.~\ref{fig::dimvsr2}) such that the rate of increase in $R^2$ score is small beyond this point. We want to understand these individual dimensions; however, they may not be orthogonal. To address this, we further orthogonalize our low-dimensional representations using PCA to ensure that each dimension encodes unique information. We then take the top few dimensions which explain most of the variance for each dataset. This contrasts with the use of PCA above to produce a baseline reduced feature set in that it is performed after supervised dimensionality reduction.

\begin{figure*}[!t]
\begin{center}
\includegraphics[width=1.0\linewidth, trim=6mm 8mm 6mm 6mm, clip=true]{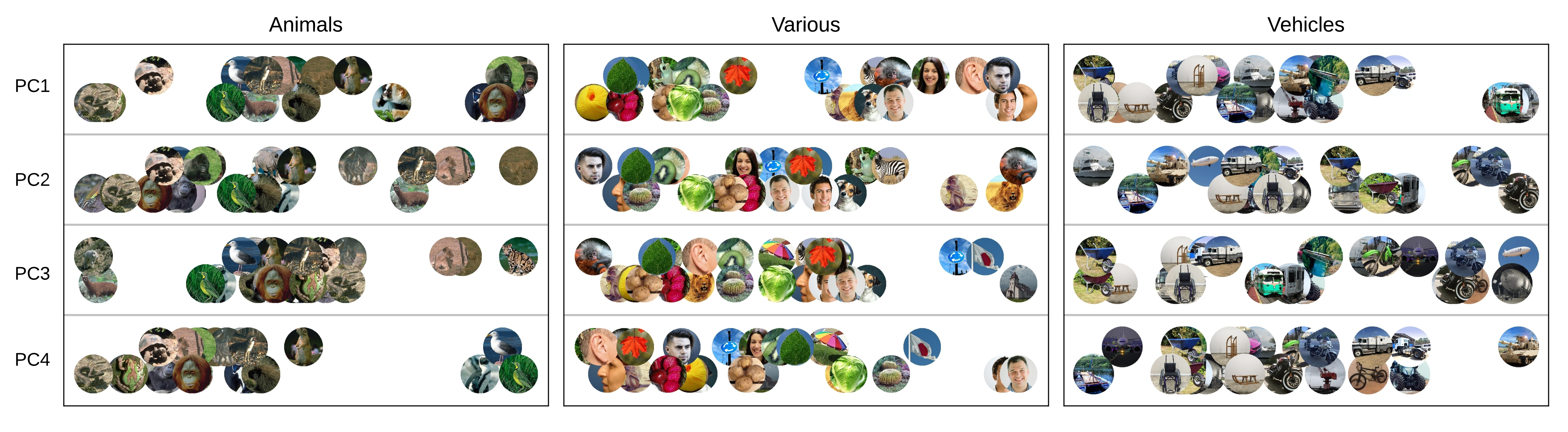}
\end{center}
\caption{Image embeddings along the top four principal components of low-dimensional SimDR representations.}
\label{fig::pca_1d}
\end{figure*}

\begin{figure*}[!t]
\begin{center}
\includegraphics[width=1.0\linewidth, trim=7mm 7mm 14mm 6mm, clip=true]{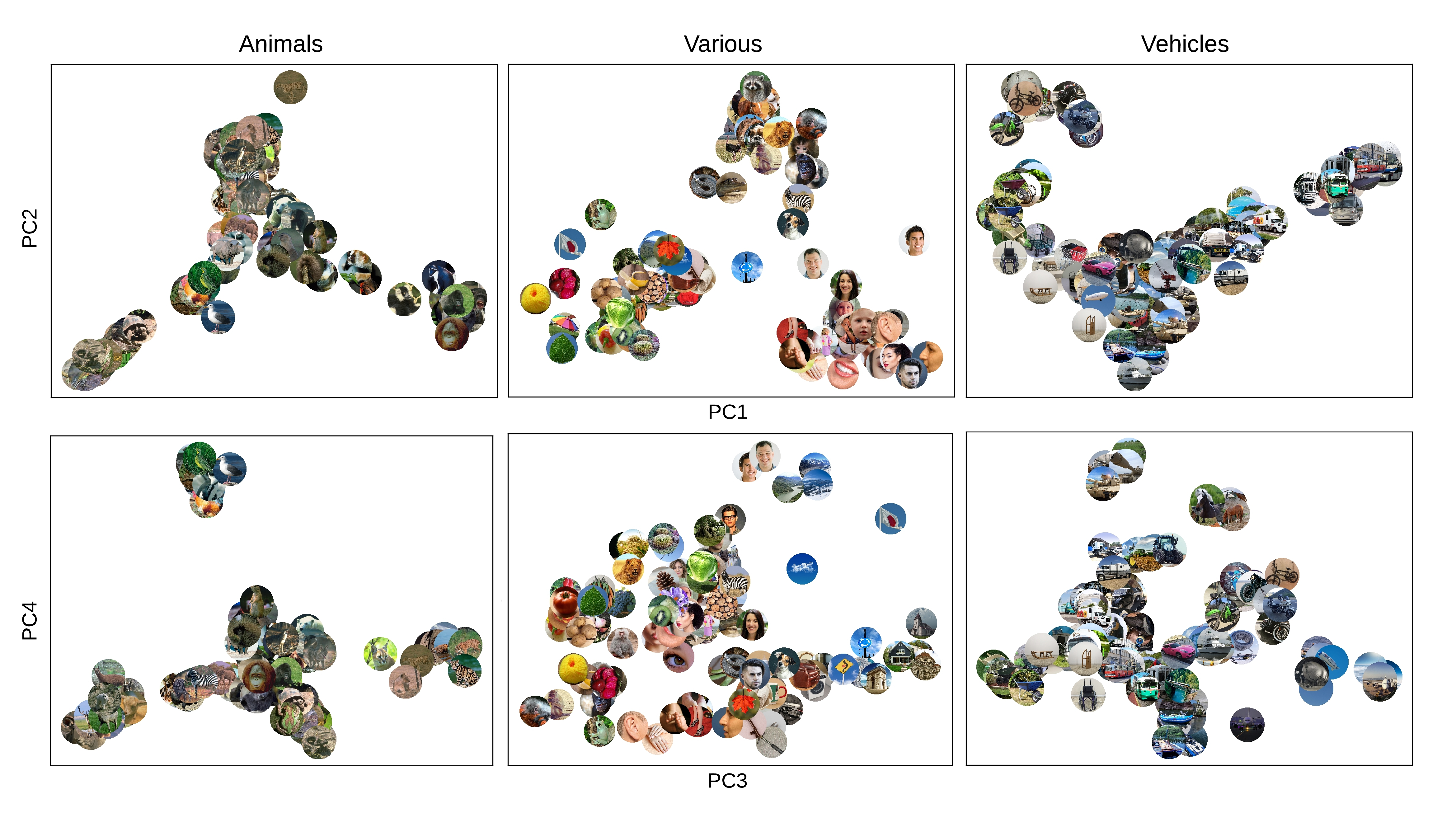}
\end{center}
\caption{Examples of image embeddings for three datasets using the top principal components of the SimDR representations.}
\label{fig::pca_2d}
\end{figure*}



\begin{figure*}[!t]
\begin{center}
\includegraphics[width=1.0\linewidth, trim=3mm 2mm 3mm 2mm, clip=true]{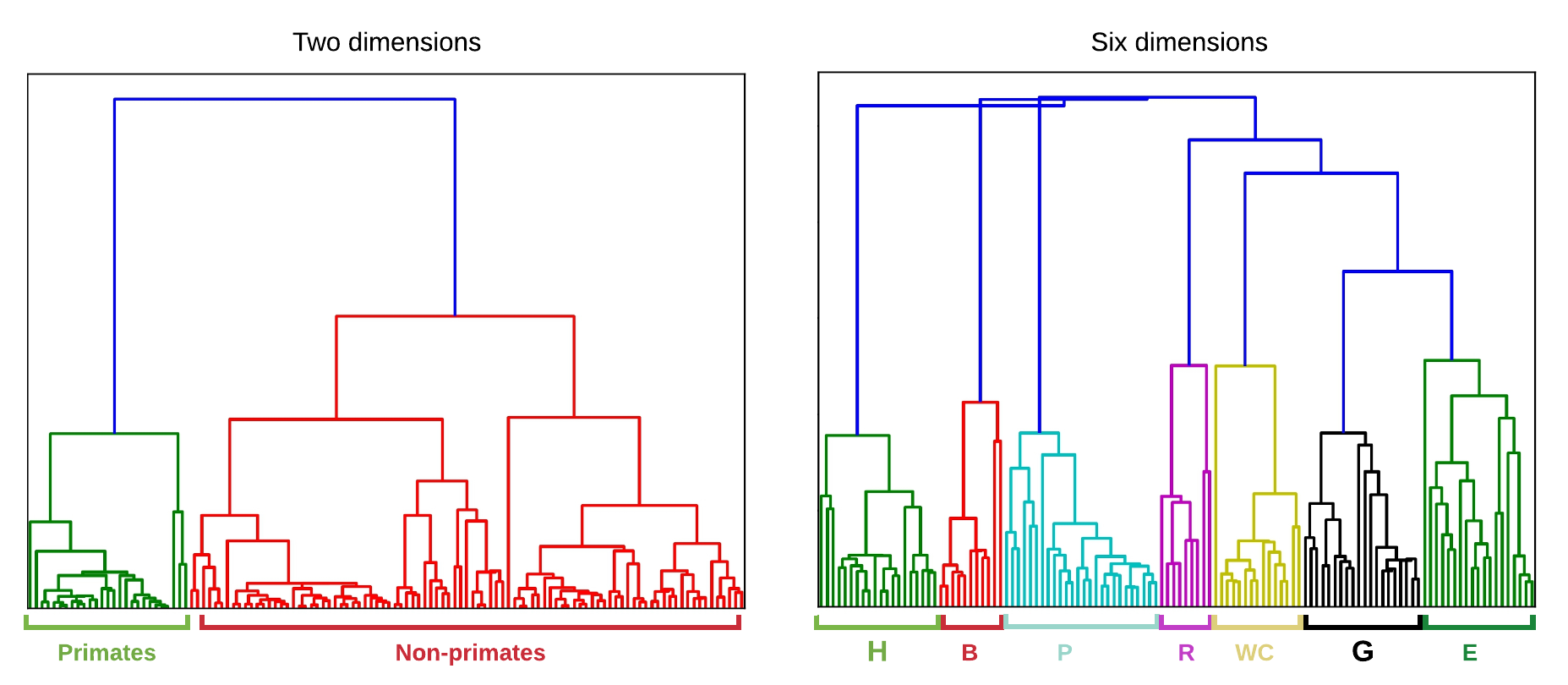}
\end{center}
\caption{Dendrograms of hierarchical clustering for 2-dimensional representations and 6-dimensional representations on \textit{animals} dataset. \textit{H: Herps, B: Birds, P: Primates, R: Rodents, WC: Wild cats, G: Grazers, E: Dogs, Bears and Large animals}.} 
\label{fig::dendrograms}
\end{figure*}

\subsection{Visualizing individual dimensions}

The ability of the low-dimensional representations to predict similarity indicates that they are efficiently encoding information essential for making similarity judgments. Hence, they can further be leveraged to understand what factors allow them to predict similarity judgments. To this end, for each of the three datasets, we visualize image embeddings along the top four principal dimensions of the low-dimensional features learned via SimDR. We visualize validation images for a single fold (out of the 6 cross validation folds), though we observe that the dimensions were consistent across all folds in terms of capturing the same concepts (Fig.~\ref{fig::pca_1d}).

We observe that the first dimension for each dataset appears to be largely continuous, and captures broad categories.
In the animals dataset, this dimension goes from non-mammals to mammals. 
The first dimension of the \textit{various} dataset goes from inanimate objects to dogs and humans. The first dimension of the \textit{vehicles} dataset shows a gradation from vehicles with two or no wheels (\textit{e.g.,} sled, wheelchair) to those with four wheels (\textit{e.g.,} trucks, buses), though the interpretation in this case is not as evident, which may stem from the low variance (12\%) captured by the top component. Some of the other principal components are also apparently interpretable and interesting. For example, the second principal component of the \textit{vehicles} dataset distinguishes water transport from land transport, the third principal component of the \textit{various} dataset distinguishes natural things from artificial ones, while the fourth dimension in the \textit{animals} dataset distinguishes birds from non-birds. Each of these individual dimensions captures a different taxonomic relationship, suggesting that such relationships are important factors in determining similarity judgments of natural images.

\subsection{Clusters formed by pairs of dimensions}
\label{subsec::clusters}
As an alternative visualization strategy, we explore 2D projections of the image representations along two of the top four principal components in Fig.~\ref{fig::pca_2d}. 
These plots are useful in observing clusters of images formed by a combination of principal components, where each cluster tells us what kind of images are considered similar by the model.

Echoing \citeA{Peterson2017}, we observe clusters for herptiles, primates, birds, wild cats, rodents, and grazers in the \textit{animals} dataset.
We see clusters for human faces and body parts, animals, vegetables, houses, and natural things in the \textit{various} dataset. 
The \textit{vehicles} dataset shows distinct clusters for trains, bikes, horses, airplanes, and tanks.

\subsection{Hierarchical similarity and bottleneck effects}


Next, we  analyze the effect of changing the width of the bottleneck layer.
We know that increasing the width improves prediction performance. Here, we are interested in interpreting the information captured by different bottleneck sizes.

To visualize this, we explore dendrograms \cite{Shepard390} for the \textit{animals} dataset. Fig.~\ref{fig::dendrograms} shows that when the size of the bottleneck layer in SimDR is 2, two clusters---primates and non-primates---are formed. 
 This suggests that belonging to the primate group is the most important trait influencing similarity judgments in the \textit{animals} dataset, which is encoded in as little as two dimensions.
At a bottleneck size of 6, however, further hierarchical structure can be seen where many more categories are present. At intermediate sizes between 2 and 6, additional clusters continue to emerge (not shown). The hierarchical structure formed by the 6-dimensional representations is closely related to that formed using human similarity data in \citeA{Peterson2017}. 

 We observe that increasing the bottleneck width introduces further categorical distinctions in other datasets too. 
For the \textit{various} dataset, at a bottleneck width of 4, we observe distinct clusters for animals and humans (and their body parts). In the case of the \textit{vehicles} dataset, 4-dimensional bottleneck layer representations preserve distinctions based on wheels. Hence, these are primary traits influencing similarity judgments which are captured at small bottleneck widths.
These results motivate a hierarchical organization of factors underlying human similarity judgments in our model, providing empirical results consistent with mathematical theories of hierarchical semantic knowledge organisation in neural networks \cite{Saxe2019}.


\section{Shared features across domains}
We have seen that each of the six individual SimDR models can discover low-dimensional representations which are predictive of similarity judgments separately for each domain. A natural question that follows from this is whether the dimensions learned by these models trained on specific domains are also shared across domains. Translating this into the framework of human judgments, the question we pose is the following: do different domains share factors underlying human similarity judgments?

\subsection{Canonical Correlation Analysis}

\begin{figure}[h]
    \centering
    \includegraphics[width=1.0\linewidth, trim=0mm 5mm 22mm 15mm, clip=true]{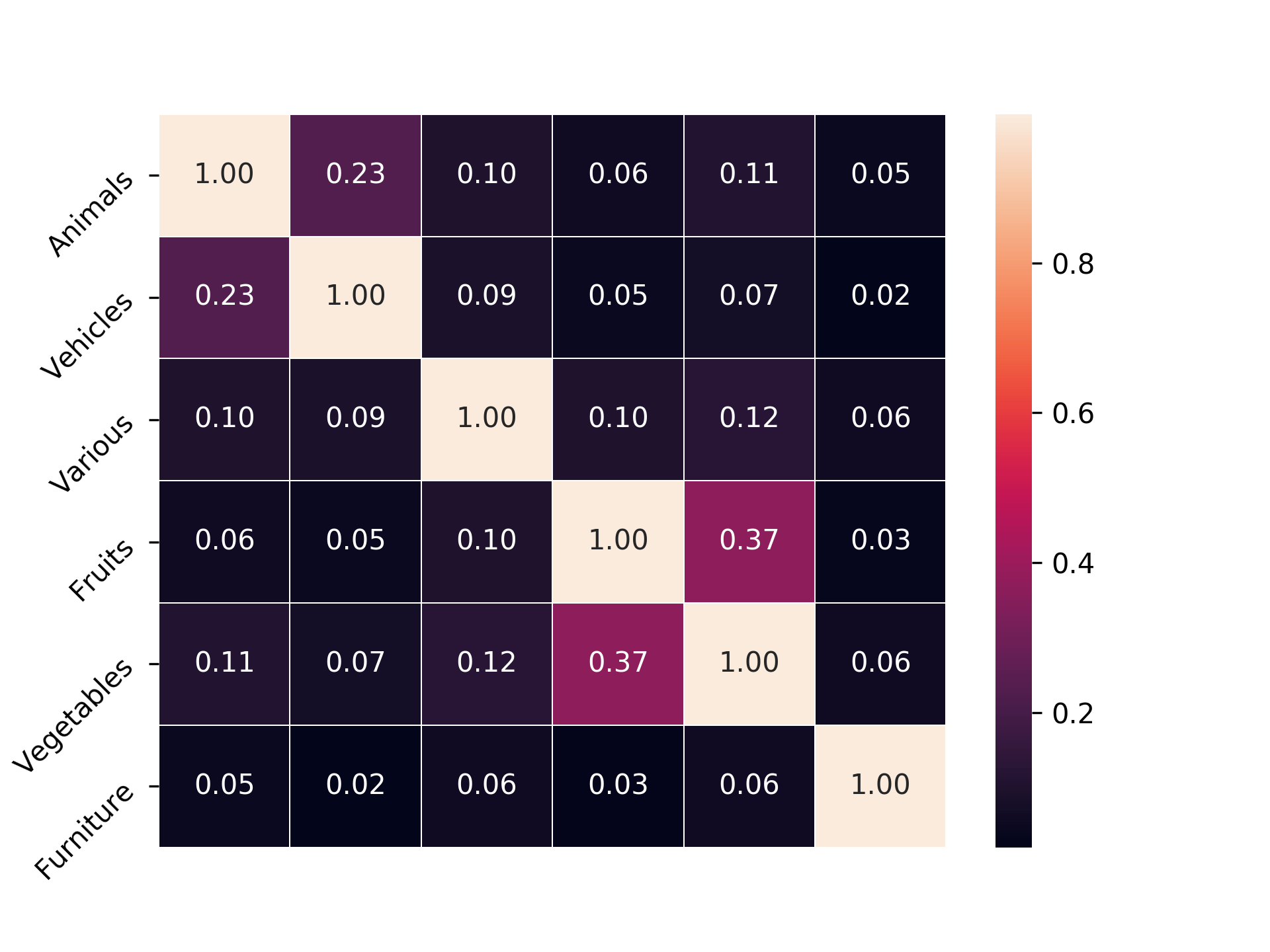}
    \caption{Inter-domain relatedness ($R^2$) as measured by regularized CCA between all domain pairs.}
    \label{fig:cca}
\end{figure}

We use L2-regularized canonical correlation analysis (CCA; \citeNP{Bilenko2016}) to evaluate the degree of shared information or factors between low-dimensional representations belonging to any two domains. From each of the six models trained on individual domains, we obtain 64-dimensional representations for all pairs of images (from all 6 datasets). We then perform regularized CCA on 64-dimensional representations from every pair of domains. 

We observe in Fig.~\ref{fig:cca} that the $R^2$ score is highest for \textit{fruits} and \textit{vegetables}, followed by \textit{animals} and \textit{vehicles}. This implies that the model trained on fruits and the model trained on vegetables have overlapping latent factors and hence, their similarity predictions are also based on some common factors. The same is true for \textit{animals} and \textit{vehicles} datasets. While it seems reasonable for \textit{fruits} and \textit{vegetables} to share common factors for similarity, the relationship between \textit{animals} and \textit{vehicles} is less clear, although we suspect it may have something to do with common backgrounds (which often contain scene information such as grass, sky, and water, unlike our other categories). 

\subsection{Domain-agnostic SimDR}

\begin{figure}[!b]
    \centering
    \includegraphics[width=1.0\linewidth, trim=6mm 2mm 16mm 14mm, clip=true]{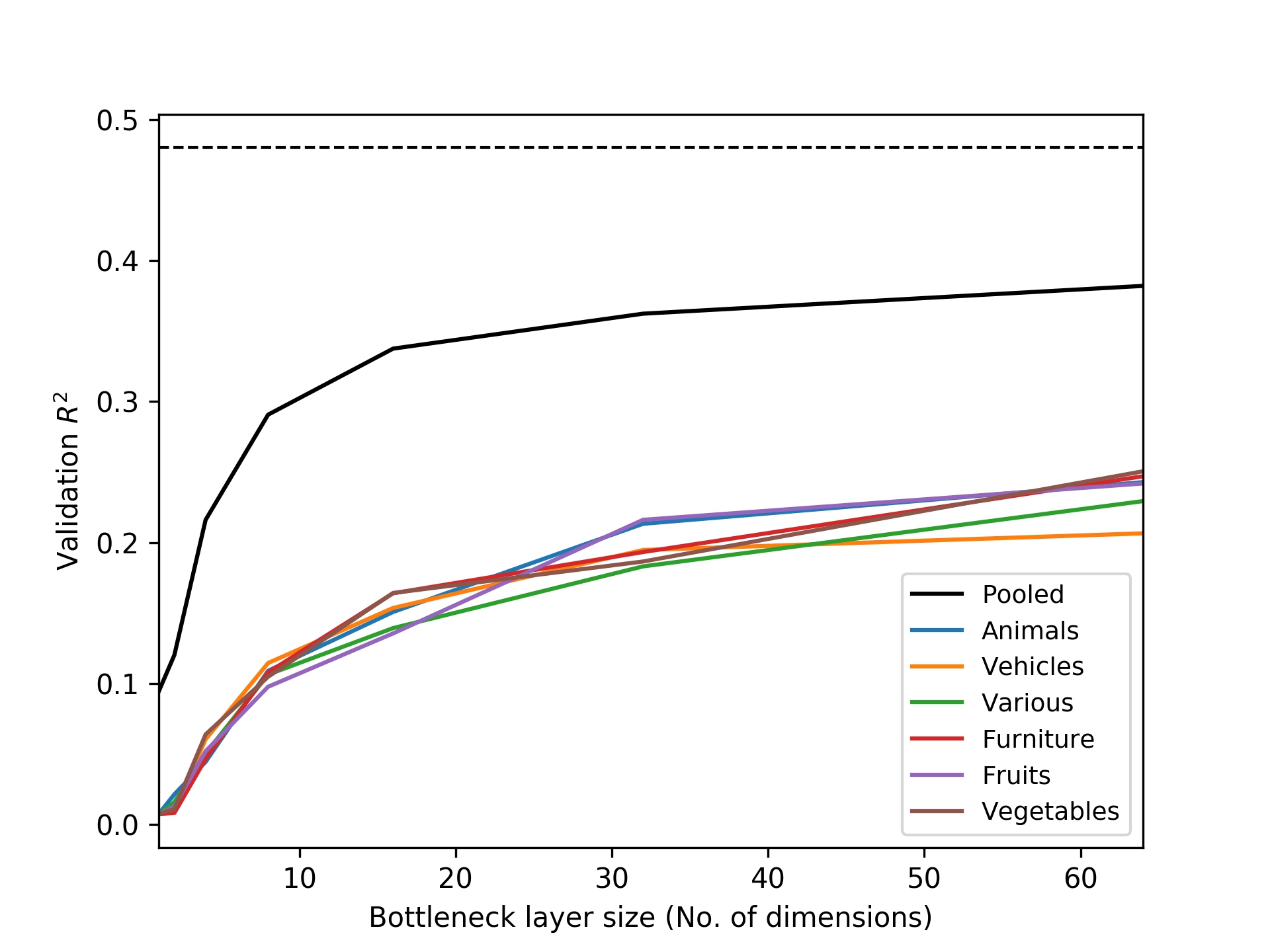}
    \caption{Performance of models tested on all domains (with varying bottleneck layer size). The dashed line shows the performance of the model trained on all domains in \citeNP{Peterson2017}. Solid lines correspond to models trained on different datasets.}
    \label{fig:pooled}
\end{figure}
\begin{figure}[!b]
    \centering
    \includegraphics[width=1.0\linewidth, trim=6mm 2mm 16mm 14mm, clip=true]{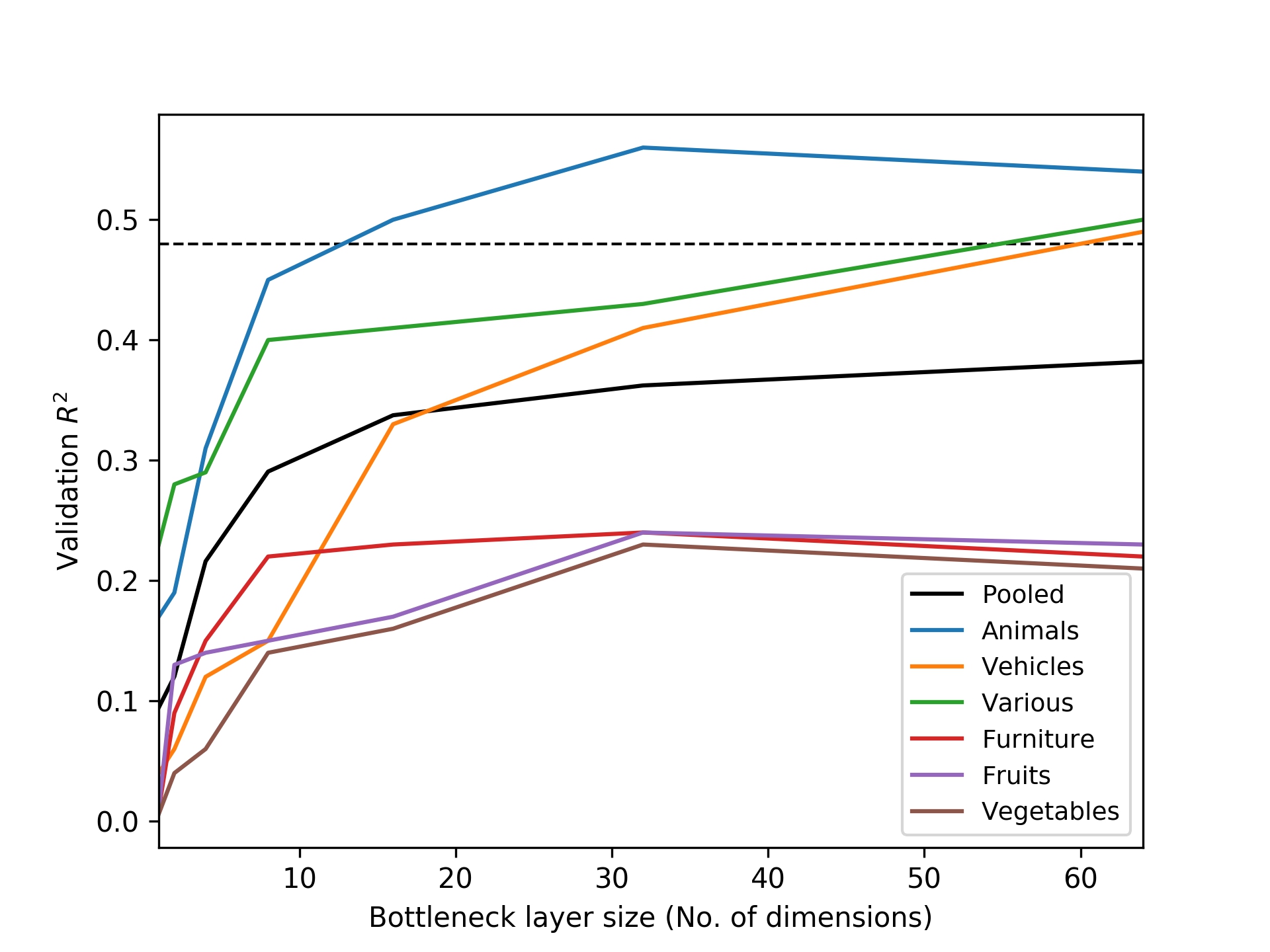}
    \caption{Performance of pooled model tested on individual domains and on all domains (with varying bottleneck layer size). The dashed line shows the performance of the model trained on all domains in \citeNP{Peterson2017}. Solid lines correspond to the pooled model tested on different datasets.}
    \label{fig:pooledtest}
\end{figure}

To determine whether a more general set of dimensions could be learned that generalizes across domains, we trained a SimDR model on image pairs from all six datasets using 6-fold cross-validation. We compared this to models trained on individual domains and tested on all others to assess how they generalize on their own.
The results, shown in Fig.~\ref{fig:pooled}, reveal that the pooled model nears saturation at a few hidden dimensions. Hence, even with a diverse dataset, few dimensions are enough to predict similarity judgments. Next, we see that the domain-specific models do not generalize well when tested on all datasets, lending credibility to our earlier claim that these models learn dimensions which are specific to individual domains. Lastly, Fig.~\ref{fig:pooledtest} shows the performance of the pooled model in predicting individual domains, and reveals that certain domains (\textit{animals}, \textit{vehicles}, \textit{various}) are well-explained by general features learned from the pool of all domains, while others require more domain-specific features (\textit{vegetables}, \textit{fruits}, \textit{furniture}). 

\section{Conclusion}

Our work shows that CNN representations can be transformed to lower dimensions---where interpretation is far less cumbersome---while still being predictive of similarity judgments. We also observe that only a few dimensions are required to predict psychological representations as opposed to a considerably larger, full set of CNN features. This finding is interesting because the deep feature sets increasingly being used in both cognitive modeling \cite<for a review, see>{ma2020neural} and neuroscience \cite{Kriegeskorte2015,kietzmann2018deep,Cichy2019} are much higher-dimensional. Indeed, some work may already suggest that our findings could generalize to modeling neural activity as well \cite{mur2013human}, though future work must bear this out.

Moreover, in this low-dimensional space, we are able to visualize individual dimensions and show that they code for unique concepts. Hence, they provide insights into potential factors that influence human similarity judgments, and potentially various other visual tasks. 
We observe that increasing the size of the bottleneck layer introduces finer levels of distinction, mirroring hierarchical clustering in human cognition. These results together show the ability of CNN representations to both predict and explain human similarity judgments using a few dimensions. 

This work takes a step towards showing that psychological representations can be predicted by far fewer dimensions than used in CNNs; and that they are not only quantitatively predictive of human similarity judgments but provide insights about how people make similarity judgments. We think our approach can help bridge the interpretation gap between CNN representations and psychological representations by providing interpretable factors which influence human similarity judgment.


\section{Acknowledgements}
This work was supported by the National Science Foundation (grant number 1718550), and the School of Engineering and Applied Sciences at Princeton University.

\bibliographystyle{apacite}

\setlength{\bibleftmargin}{.125in}
\setlength{\bibindent}{-\bibleftmargin}

\bibliography{refs}

\begin{thebibliography}{}

\bibitem [\protect \citeauthoryear {%
Bilenko%
\ \BBA {} Gallant%
}{%
Bilenko%
\ \BBA {} Gallant%
}{%
{\protect \APACyear {2016}}%
}]{%
Bilenko2016}
\APACinsertmetastar {%
Bilenko2016}%
\begin{APACrefauthors}%
Bilenko, N\BPBI Y.%
\BCBT {}\ \BBA {} Gallant, J\BPBI L.%
\end{APACrefauthors}%
\unskip\
\newblock
\APACrefYearMonthDay{2016}{}{}.
\newblock
{\BBOQ}\APACrefatitle {Pyrcca: Regularized Kernel Canonical Correlation
  Analysis in {P}ython and Its Applications to Neuroimaging} {Pyrcca:
  Regularized kernel canonical correlation analysis in {P}ython and its
  applications to neuroimaging}.{\BBCQ}
\newblock
\APACjournalVolNumPages{Frontiers in Neuroinformatics}{10}{}{49}.
\newblock
\begin{APACrefDOI} \doi{10.3389/fninf.2016.00049} \end{APACrefDOI}
\PrintBackRefs{\CurrentBib}

\bibitem [\protect \citeauthoryear {%
Cichy%
\ \BBA {} Kaiser%
}{%
Cichy%
\ \BBA {} Kaiser%
}{%
{\protect \APACyear {2019}}%
}]{%
Cichy2019}
\APACinsertmetastar {%
Cichy2019}%
\begin{APACrefauthors}%
Cichy, R\BPBI M.%
\BCBT {}\ \BBA {} Kaiser, D.%
\end{APACrefauthors}%
\unskip\
\newblock
\APACrefYearMonthDay{2019}{}{}.
\newblock
{\BBOQ}\APACrefatitle {Deep Neural Networks as Scientific Models} {Deep neural
  networks as scientific models}.{\BBCQ}
\newblock
\APACjournalVolNumPages{Trends in Cognitive Sciences}{23}{4}{305 - 317}.
\newblock
\begin{APACrefDOI} \doi{https://doi.org/10.1016/j.tics.2019.01.009}
  \end{APACrefDOI}
\PrintBackRefs{\CurrentBib}

\bibitem [\protect \citeauthoryear {%
Cohen%
}{%
Cohen%
}{%
{\protect \APACyear {2000}}%
}]{%
Cohen2000}
\APACinsertmetastar {%
Cohen2000}%
\begin{APACrefauthors}%
Cohen, G.%
\end{APACrefauthors}%
\unskip\
\newblock
\APACrefYearMonthDay{2000}{03}{}.
\newblock
{\BBOQ}\APACrefatitle {Hierarchical models in cognition: Do they have
  psychological reality?} {Hierarchical models in cognition: Do they have
  psychological reality?}{\BBCQ}
\newblock
\APACjournalVolNumPages{European Journal of Cognitive Psychology}{12}{}{1-36}.
\newblock
\begin{APACrefDOI} \doi{10.1080/095414400382181} \end{APACrefDOI}
\PrintBackRefs{\CurrentBib}

\bibitem [\protect \citeauthoryear {%
Diesendruck%
\ \BBA {} Bloom%
}{%
Diesendruck%
\ \BBA {} Bloom%
}{%
{\protect \APACyear {2003}}%
}]{%
Diesendruck2003}
\APACinsertmetastar {%
Diesendruck2003}%
\begin{APACrefauthors}%
Diesendruck, G.%
\BCBT {}\ \BBA {} Bloom, P.%
\end{APACrefauthors}%
\unskip\
\newblock
\APACrefYearMonthDay{2003}{}{}.
\newblock
{\BBOQ}\APACrefatitle {How Specific is the Shape Bias?} {How specific is the
  shape bias?}{\BBCQ}
\newblock
\APACjournalVolNumPages{Child Development}{74}{1}{168-178}.
\PrintBackRefs{\CurrentBib}

\bibitem [\protect \citeauthoryear {%
Kietzmann%
, McClure%
\BCBL {}\ \BBA {} Kriegeskorte%
}{%
Kietzmann%
\ \protect \BOthers {.}}{%
{\protect \APACyear {2019}}%
}]{%
kietzmann2018deep}
\APACinsertmetastar {%
kietzmann2018deep}%
\begin{APACrefauthors}%
Kietzmann, T\BPBI C.%
, McClure, P.%
\BCBL {}\ \BBA {} Kriegeskorte, N.%
\end{APACrefauthors}%
\unskip\
\newblock
\APACrefYearMonthDay{2019}{01}{}.
\newblock
{\BBOQ}\APACrefatitle {Deep Neural Networks in Computational Neuroscience}
  {Deep neural networks in computational neuroscience}.{\BBCQ}
\newblock
\APACjournalVolNumPages{Oxford Research Encyclopedia of Neuroscience}{}{}{}.
\PrintBackRefs{\CurrentBib}

\bibitem [\protect \citeauthoryear {%
Kriegeskorte%
}{%
Kriegeskorte%
}{%
{\protect \APACyear {2015}}%
}]{%
Kriegeskorte2015}
\APACinsertmetastar {%
Kriegeskorte2015}%
\begin{APACrefauthors}%
Kriegeskorte, N.%
\end{APACrefauthors}%
\unskip\
\newblock
\APACrefYearMonthDay{2015}{}{}.
\newblock
{\BBOQ}\APACrefatitle {Deep Neural Networks: A New Framework for Modeling
  Biological Vision and Brain Information Processing} {Deep neural networks: A
  new framework for modeling biological vision and brain information
  processing}.{\BBCQ}
\newblock
\APACjournalVolNumPages{Annual Review of Vision Science}{1}{1}{417-446}.
\PrintBackRefs{\CurrentBib}

\bibitem [\protect \citeauthoryear {%
Kubilius%
, Bracci%
\BCBL {}\ \BBA {} Op~de Beeck%
}{%
Kubilius%
\ \protect \BOthers {.}}{%
{\protect \APACyear {2016}}%
}]{%
Kubilius2016}
\APACinsertmetastar {%
Kubilius2016}%
\begin{APACrefauthors}%
Kubilius, J.%
, Bracci, S.%
\BCBL {}\ \BBA {} Op~de Beeck, H\BPBI P.%
\end{APACrefauthors}%
\unskip\
\newblock
\APACrefYearMonthDay{2016}{04}{}.
\newblock
{\BBOQ}\APACrefatitle {Deep Neural Networks as a Computational Model for Human
  Shape Sensitivity} {Deep neural networks as a computational model for human
  shape sensitivity}.{\BBCQ}
\newblock
\APACjournalVolNumPages{PLOS Computational Biology}{12}{4}{1-26}.
\PrintBackRefs{\CurrentBib}

\bibitem [\protect \citeauthoryear {%
Lake%
, Zaremba%
, Fergus%
\BCBL {}\ \BBA {} Gureckis%
}{%
Lake%
\ \protect \BOthers {.}}{%
{\protect \APACyear {2015}}%
}]{%
Lake2015}
\APACinsertmetastar {%
Lake2015}%
\begin{APACrefauthors}%
Lake, B.%
, Zaremba, W.%
, Fergus, R.%
\BCBL {}\ \BBA {} Gureckis, T.%
\end{APACrefauthors}%
\unskip\
\newblock
\APACrefYearMonthDay{2015}{}{}.
\newblock
{\BBOQ}\APACrefatitle {Deep neural networks predict category typicality ratings
  for images} {Deep neural networks predict category typicality ratings for
  images}.{\BBCQ}
\newblock
\BIn{} R.~Dale\ \BOthers {.}\ (\BEDS), \APACrefbtitle {Proceedings of the 37th
  {A}nnual {C}onference of the {C}ognitive {S}cience {S}ociety.} {Proceedings
  of the 37th {A}nnual {C}onference of the {C}ognitive {S}cience {S}ociety.}
\PrintBackRefs{\CurrentBib}

\bibitem [\protect \citeauthoryear {%
Ma%
\ \BBA {} Peters%
}{%
Ma%
\ \BBA {} Peters%
}{%
{\protect \APACyear {2020}}%
}]{%
ma2020neural}
\APACinsertmetastar {%
ma2020neural}%
\begin{APACrefauthors}%
Ma, W\BPBI J.%
\BCBT {}\ \BBA {} Peters, B.%
\end{APACrefauthors}%
\unskip\
\newblock
\APACrefYearMonthDay{2020}{}{}.
\newblock
{\BBOQ}\APACrefatitle {A neural network walks into a lab: towards using deep
  nets as models for human behavior} {A neural network walks into a lab:
  towards using deep nets as models for human behavior}.{\BBCQ}
\newblock
\APACjournalVolNumPages{arXiv preprint arXiv:2005.02181}{}{}{}.
\PrintBackRefs{\CurrentBib}

\bibitem [\protect \citeauthoryear {%
Medin%
, Goldstone%
\BCBL {}\ \BBA {} Gentner%
}{%
Medin%
\ \protect \BOthers {.}}{%
{\protect \APACyear {1993}}%
}]{%
Medin1993}
\APACinsertmetastar {%
Medin1993}%
\begin{APACrefauthors}%
Medin, D.%
, Goldstone, R.%
\BCBL {}\ \BBA {} Gentner, D.%
\end{APACrefauthors}%
\unskip\
\newblock
\APACrefYearMonthDay{1993}{04}{}.
\newblock
{\BBOQ}\APACrefatitle {Respects for Similarity} {Respects for
  similarity}.{\BBCQ}
\newblock
\APACjournalVolNumPages{Psychological Review}{100}{}{254-278}.
\newblock
\begin{APACrefDOI} \doi{10.1037/0033-295X.100.2.254} \end{APACrefDOI}
\PrintBackRefs{\CurrentBib}

\bibitem [\protect \citeauthoryear {%
Mur%
\ \protect \BOthers {.}}{%
Mur%
\ \protect \BOthers {.}}{%
{\protect \APACyear {2013}}%
}]{%
mur2013human}
\APACinsertmetastar {%
mur2013human}%
\begin{APACrefauthors}%
Mur, M.%
, Meys, M.%
, Bodurka, J.%
, Goebel, R.%
, Bandettini, P\BPBI A.%
\BCBL {}\ \BBA {} Kriegeskorte, N.%
\end{APACrefauthors}%
\unskip\
\newblock
\APACrefYearMonthDay{2013}{}{}.
\newblock
{\BBOQ}\APACrefatitle {Human object-similarity judgments reflect and transcend
  the primate-IT object representation} {Human object-similarity judgments
  reflect and transcend the primate-it object representation}.{\BBCQ}
\newblock
\APACjournalVolNumPages{Frontiers in {P}sychology}{4}{}{128}.
\PrintBackRefs{\CurrentBib}

\bibitem [\protect \citeauthoryear {%
Peterson%
, Abbott%
\BCBL {}\ \BBA {} Griffiths%
}{%
Peterson%
\ \protect \BOthers {.}}{%
{\protect \APACyear {2018}}%
}]{%
Peterson2017}
\APACinsertmetastar {%
Peterson2017}%
\begin{APACrefauthors}%
Peterson, J\BPBI C.%
, Abbott, J\BPBI T.%
\BCBL {}\ \BBA {} Griffiths, T\BPBI L.%
\end{APACrefauthors}%
\unskip\
\newblock
\APACrefYearMonthDay{2018}{}{}.
\newblock
{\BBOQ}\APACrefatitle {Evaluating (and Improving) the Correspondence Between
  Deep Neural Networks and Human Representations} {Evaluating (and improving)
  the correspondence between deep neural networks and human
  representations}.{\BBCQ}
\newblock
\APACjournalVolNumPages{Cognitive Science}{42}{8}{2648--2669}.
\PrintBackRefs{\CurrentBib}

\bibitem [\protect \citeauthoryear {%
Rogers%
\ \BBA {} McClelland%
}{%
Rogers%
\ \BBA {} McClelland%
}{%
{\protect \APACyear {2004}}%
}]{%
Rogers2004SemanticCA}
\APACinsertmetastar {%
Rogers2004SemanticCA}%
\begin{APACrefauthors}%
Rogers, T\BPBI T.%
\BCBT {}\ \BBA {} McClelland, J\BPBI L.%
\end{APACrefauthors}%
\unskip\
\newblock
\APACrefYearMonthDay{2004}{}{}.
\newblock
{\BBOQ}\APACrefatitle {Semantic Cognition: A Parallel Distributed Processing
  Approach} {Semantic cognition: A parallel distributed processing
  approach}.{\BBCQ}.
\PrintBackRefs{\CurrentBib}

\bibitem [\protect \citeauthoryear {%
Rumelhart%
\ \BBA {} Todd%
}{%
Rumelhart%
\ \BBA {} Todd%
}{%
{\protect \APACyear {1993}}%
}]{%
Rumelhart1993}
\APACinsertmetastar {%
Rumelhart1993}%
\begin{APACrefauthors}%
Rumelhart, D\BPBI E.%
\BCBT {}\ \BBA {} Todd, P\BPBI M.%
\end{APACrefauthors}%
\unskip\
\newblock
\APACrefYearMonthDay{1993}{}{}.
\newblock
{\BBOQ}\APACrefatitle {Learning and Connectionist Representations} {Learning
  and connectionist representations}.{\BBCQ}
\newblock
\BIn{} \APACrefbtitle {Attention and Performance XIV (Silver Jubilee Volume):
  Synergies in Experimental Psychology, Artificial Intelligence, and Cognitive
  Neuroscience} {Attention and performance xiv (silver jubilee volume):
  Synergies in experimental psychology, artificial intelligence, and cognitive
  neuroscience}\ (\BPG~3–30).
\newblock
\APACaddressPublisher{Cambridge, MA, USA}{MIT Press}.
\PrintBackRefs{\CurrentBib}

\bibitem [\protect \citeauthoryear {%
Sanders%
\ \BBA {} Nosofsky%
}{%
Sanders%
\ \BBA {} Nosofsky%
}{%
{\protect \APACyear {2020}}%
}]{%
Sanders2020}
\APACinsertmetastar {%
Sanders2020}%
\begin{APACrefauthors}%
Sanders, C\BPBI A.%
\BCBT {}\ \BBA {} Nosofsky, R\BPBI M.%
\end{APACrefauthors}%
\unskip\
\newblock
\APACrefYearMonthDay{2020}{}{}.
\newblock
{\BBOQ}\APACrefatitle {Training Deep Networks to Construct a Psychological
  Feature Space for a Natural-Object Category Domain} {Training deep networks
  to construct a psychological feature space for a natural-object category
  domain}.{\BBCQ}
\newblock
\APACjournalVolNumPages{Computational Brain \& Behavior}{}{}{}.
\newblock
\begin{APACrefDOI} \doi{https://doi.org/10.1007/s42113-020-00073-z}
  \end{APACrefDOI}
\PrintBackRefs{\CurrentBib}

\bibitem [\protect \citeauthoryear {%
Saxe%
, McClelland%
\BCBL {}\ \BBA {} Ganguli%
}{%
Saxe%
\ \protect \BOthers {.}}{%
{\protect \APACyear {2019}}%
}]{%
Saxe2019}
\APACinsertmetastar {%
Saxe2019}%
\begin{APACrefauthors}%
Saxe, A\BPBI M.%
, McClelland, J\BPBI L.%
\BCBL {}\ \BBA {} Ganguli, S.%
\end{APACrefauthors}%
\unskip\
\newblock
\APACrefYearMonthDay{2019}{}{}.
\newblock
{\BBOQ}\APACrefatitle {A mathematical theory of semantic development in deep
  neural networks} {A mathematical theory of semantic development in deep
  neural networks}.{\BBCQ}
\newblock
\APACjournalVolNumPages{Proceedings of the National Academy of
  Sciences}{116}{23}{11537--11546}.
\newblock
\begin{APACrefDOI} \doi{10.1073/pnas.1820226116} \end{APACrefDOI}
\PrintBackRefs{\CurrentBib}

\bibitem [\protect \citeauthoryear {%
Shepard%
}{%
Shepard%
}{%
{\protect \APACyear {1980}}%
}]{%
Shepard390}
\APACinsertmetastar {%
Shepard390}%
\begin{APACrefauthors}%
Shepard, R\BPBI N.%
\end{APACrefauthors}%
\unskip\
\newblock
\APACrefYearMonthDay{1980}{}{}.
\newblock
{\BBOQ}\APACrefatitle {Multidimensional Scaling, Tree-Fitting, and Clustering}
  {Multidimensional scaling, tree-fitting, and clustering}.{\BBCQ}
\newblock
\APACjournalVolNumPages{Science}{210}{4468}{390--398}.
\PrintBackRefs{\CurrentBib}

\bibitem [\protect \citeauthoryear {%
Simonyan%
\ \BBA {} Zisserman%
}{%
Simonyan%
\ \BBA {} Zisserman%
}{%
{\protect \APACyear {2015}}%
}]{%
Simonyan15}
\APACinsertmetastar {%
Simonyan15}%
\begin{APACrefauthors}%
Simonyan, K.%
\BCBT {}\ \BBA {} Zisserman, A.%
\end{APACrefauthors}%
\unskip\
\newblock
\APACrefYearMonthDay{2015}{}{}.
\newblock
{\BBOQ}\APACrefatitle {Very Deep Convolutional Networks for Large-Scale Image
  Recognition} {Very deep convolutional networks for large-scale image
  recognition}.{\BBCQ}
\newblock
\BIn{} \APACrefbtitle {{I}nternational {C}onference on {L}earning
  {R}epresentations.} {{I}nternational {C}onference on {L}earning
  {R}epresentations.}
\PrintBackRefs{\CurrentBib}

\end{thebibliography}

\end{document}